\documentclass[usenatbib]{mnras}

\usepackage{newtxtext,newtxmath}
\usepackage{ae,aecompl}

\usepackage{graphicx}	% Including figure files
\usepackage{amsmath}	% Advanced maths commands
\usepackage{amssymb}	% Extra maths symbols
\usepackage{threeparttable}
\usepackage{ulem}
\usepackage{float}
\newcommand{\dif}{\mathrm{d}}
\defcitealias{sas+19}{Paper~I}
\newcommand{\paper}{\citetalias{sas+19}}

\title[The initial evolution of millisecond magnetars]{The initial evolution of millisecond magnetars: an analytical solution}

\author[S. {\c C}{\i}k{\i}nto{\u g}lu et al.]{
S. {\c C}{\i}k{\i}nto{\u g}lu$^{1}$\thanks{E-mail: cikintoglus@itu.edu.tr},
S. {\c S}a{\c s}maz Mu{\c s}$^{1}$,
K.~Yavuz Ek\c{s}i$^{1}$
\\
$^{1}$Istanbul Technical University, Faculty  of Science  and  Letters, Physics Engineering  Department, 34469, Istanbul, Turkey\\
}

\date{Accepted XXX. Received YYY; in original form ZZZ}

\pubyear{2019}

\begin{document}
\label{firstpage}
\pagerange{\pageref{firstpage}--\pageref{lastpage}}
\maketitle

\begin{abstract}
Millisecond magnetars are often invoked as the central engine of some gamma-ray bursts (GRBs), specifically the ones showing a plateau phase. We argue that an apparent plateau phase may not be realized if the magnetic field of the nascent magnetar is in a transient rapid decay stage. Some GRBs that lack a clear plateau phase may also be hosting millisecond magnetars.  We present an approximate analytical solution of the coupled set of equations describing the evolution of the angular velocity and the inclination angle between rotation and magnetic axis of a neutron star in the presence of a co-rotating plasma. We also show how the solution can be generalized to the case of evolving magnetic fields. We determine the evolution of the spin period, inclination angle, magnetic dipole moment and braking index of six putative magnetars associated with GRB 091018, GRB 070318, GRB 080430, GRB 090618, GRB 110715A, GRB 140206A through fitting, via Bayesian analysis, the X-ray afterglow light curves by using our recent model [{\c S}a{\c s}maz Mu{\c s} et al.\ 2019].  We find that within the first day following the formation of the millisecond magnetar, the inclination angle aligns rapidly, the magnetic dipole field may decay by a few times and the braking index varies by an order of magnitude.
\end{abstract}

\begin{keywords}
gamma-ray burst: general -- stars: magnetars
\end{keywords}

%%%%%%%%%%%%%%%%%%%%%%%%%%%%%%%%%%%%%%%%%%%%%%%%%%

%%%%%%%%%%%%%%%%% BODY OF PAPER %%%%%%%%%%%%%%%%%%

\section{Introduction}

Gamma-ray bursts (GRBs) are highly energetic explosions with durations of milliseconds to minutes \citep{Lyu+03,Zha+04,Pir05,Kum+15}. The prompt emission is followed by an X-ray afterglow \citep{cos+97}. It is considered that the central engine of some of the GRBs could be strongly magnetized rapidly spinning neutron stars, i.e.\ millisecond magnetars \citep{uso92,dun92}, specifically the ones showing a plateau stage in their afterglows \citep{dai98a,dai98b}. The spin-down power of a millisecond magnetar, $L_{\rm sd}=-I\Omega \dot{\Omega}$ where $I$ is the moment of inertia of the star, $\Omega$ is the angular velocity, and the dot denotes the derivative with respect to time, is employed for explaining the X-ray afterglows:
\begin{equation}
L_{\rm X}=\eta L_{\rm sd},
\label{eq:LX}
\end{equation}
where $\eta$ is an efficiency coefficient. In the case of spin-down under magnetic dipole torque $I\dot{\Omega}= -2\mu^2 \sin^2 \alpha \Omega^3/3c^3$ 
where $\mu$ is the magnetic dipole moment and $\alpha$ is the inclination angle between rotation and magnetic axis, an exact analytic solution, $\Omega=\Omega_0 (1+t/t_0)^{-1/2}$, can be obtained where $\Omega_0$ is the initial angular velocity and $t_0=3Ic^3/(2\mu\Omega_0\sin\alpha)^2$ is the spin-down time-scale. This leads to $L_{\rm X}=\eta L_0(1+t/t_0)^{-2}$ where $L_0=2\mu^2\Omega^4_0\sin^2\alpha/3c^3$. The spin-down time-scale $t_0$ determines the duration of the plateau phase which is followed by the rapid-decay stage $L_{\rm X} \propto t^{-2}$. This model has been generalized by \citet{las+17} to infer the braking indices of nascent magnetars \citep[see also][]{lu+19}.

%%%%%%%%%%%%%%%%%%%%%%%%%%%%%%%%%%%
%%%%%%%% FIGURE 1
%%%%%%%%%%%%%%%%%%%%%%%%%%%%%%%%%%%
\begin{figure*}
\includegraphics[width=\textwidth]{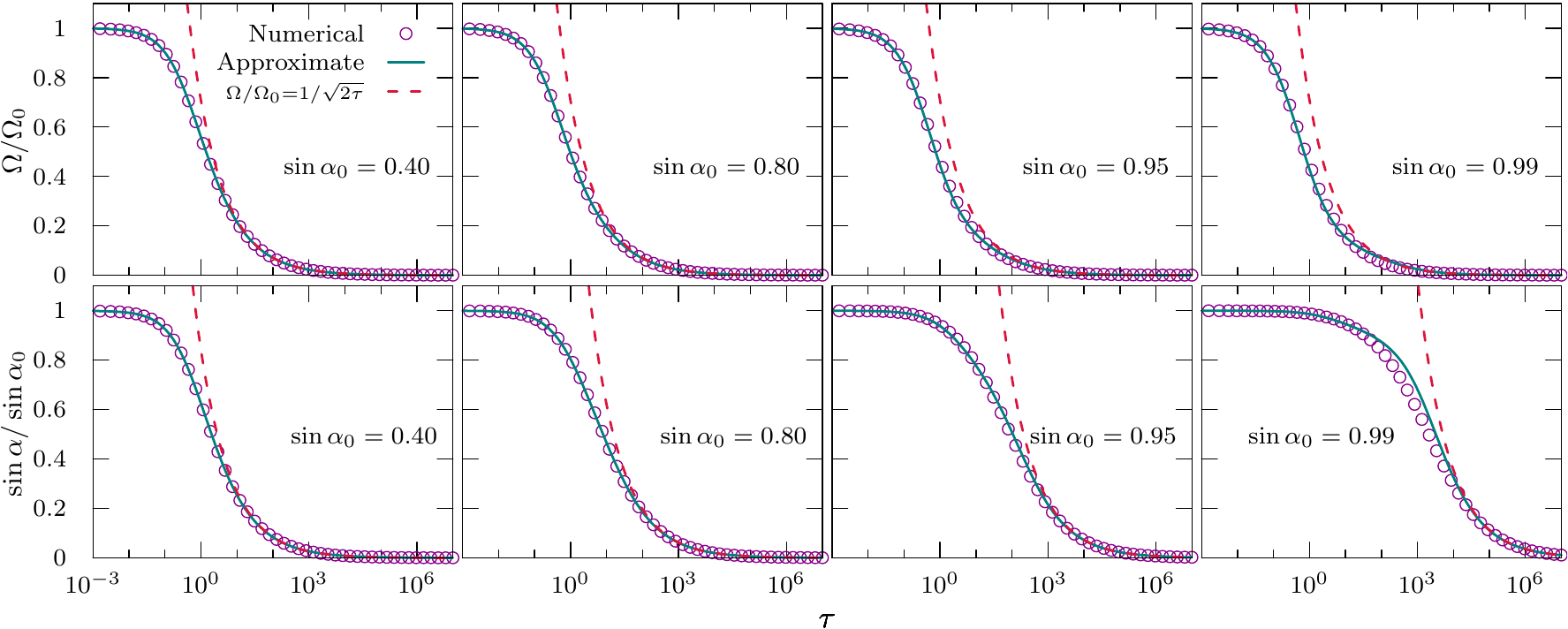}
\caption{The evolution of angular velocity {\bf (upper panels)} and inclination angle {\bf (lower panels)} for various initial angles. The circles denote the numerical solutions, the solid lines denote the approximate analytical solutions given in Equations~(\ref{eq:app_sol2})-(\ref{eq:app_sol_w}) and the dashed lines (red in electronic version) represent the late-time limits given in Equation (\ref{eq:lim2}). The time in the $x$-axis is in units of spin-down time scale defined in Eqn.~\ref{eq:t_tau}. 
\label{fig:app_sol}}
\end{figure*}

The solution given above, employed by many previous work, neglects the alignment component of the dipole torque \citep{mic70,dav70}. It also assumes a constant magnetic dipole  moment rotating in vacuum. Initially, the magnetar is far from an equilibrium stage and its just generated magnetic field may be in a rapid relaxation stage \citep{gep+06,ben+17}. Because of this rapid decay of the field, the spin-down power may decline so fast that a clear `plateau phase' may not be realized. 

In this work we employ the recent model proposed by \citet{sas+19} (hereafter \paper) for modeling the X-ray afterglows of six putative magnetars associated with GRB afterglow light curves, GRBs 091018, 070318, 080430, 090618, 110715A and 140206A. This model assumes the magnetar has a corotating plasma \citep{gol69} and employs the appropriate spin-down \citep{Spi06} and alignment \citep{phi+14a} torque components. It also assumes an exponential relaxation of the magnetic dipole moment (\paper).

We review the model equations in Section \ref{sec:modeleq}. We present an approximate analytical solution for the model equations in Section~\ref{sec:appsol}, GRB sample used in this work in Section~\ref{sec:grbsamp} and the method for fitting the model to the GRB afterglow light curves in Section~\ref{sec:parest}. We present our results in Section~\ref{sec:results} and discuss the implications of our findings in Section~\ref{sec:discuss}.

%%%%%%%%%%%%%%%%%%%%%%%%%%%%%%%%%%%%%%%%%%%%%%
%%%%%%%%%%%%%%%%%%%%%%%%%%%%%%%%%%%%%%%%%%%%%%
\section{Method}
\label{sec:method}
%%%%%%%%%%%%%%%%%%%%%%%%%%%%%%%%%%%%%%%%%%%%%%
%%%%%%%%%%%%%%%%%%%%%%%%%%%%%%%%%%%%%%%%%%%%%%

%%%%%%%%%%%%%%%%%%%%%%%%%%%%%%%%%%%%%
\subsection{Model equations}
\label{sec:modeleq}
%%%%%%%%%%%%%%%%%%%%%%%%%%%%%%%%%%%%%

We employ the model recently set-up by \paper~to fit the X-ray afterglow light curves of 7 GRBs. This model is a set of three ordinary differential equations (ODEs) which employs the spin-down \citep{Spi06}
%%%%%%%%%%%%%%%%
\begin{equation}
I \frac{{\rm d} \Omega}{{\rm d}t} = -\frac{ \mu^2 \Omega^3}{c^3} (1 + \sin^2 \alpha ), 
\label{eq:mdr1}
\end{equation}
%%%%%%%%%%%%%%%%
and alignment \citep{phi+14a}
%%%%%%%%%%%%%%%%
\begin{equation}
I \frac{{\rm d} \alpha}{{\rm d}t} = -\frac{ \mu^2 \Omega^2}{c^3} \sin \alpha \cos \alpha 
\label{eq:mdr2},
\end{equation}
%%%%%%%%%%%%%%%%
components of the magnetic dipole torque in the presence of a corotating plasma \citep{Hon+65,gol69}, and a simple prescription for the evolution of the magnetic dipole moment 
%%%%%%%%%%%%%%%%
\begin{equation}
\dot{\mu} = -(\mu-\mu_\infty)/t_{\rm m},
\label{eq:dotmu}
\end{equation}
%%%%%%%%%%%%%%%%
(\paper) where $\mu_{\infty}$ is the settling value of the magnetic dipole moment and $t_{\rm m}$ is its evolution time-scale. This model predicts the braking index to be
%%%%%%%%%%%%%%%%
\begin{equation}
n \equiv \frac{\Omega\ddot{\Omega}}{\dot{\Omega}^2}= 3+2\left[\frac{\sin\alpha\cos\alpha}{1+\sin^2\alpha}\right]^2+2\frac{\Omega}{\dot{\Omega}}\frac{\dot{\mu}}{\mu}.
\label{eq:brake}
\end{equation}
%%%%%%%%%%%%%%%%
The first two equations, Eqn.(\ref{eq:mdr1}) and Eqn.(\ref{eq:mdr2}), are coupled while Equation (\ref{eq:dotmu}) can be solved independently to give
\begin{equation}
\mu = \mu_{\infty}+\left(\mu_0-\mu_{\infty}\right) {\rm e}^{-t/t_{\rm m}},
\label{eq:mu}
\end{equation}
where $\mu_0$ is the initial magnetic dipole moment of the magnetar.

%%%%%%%%%%%%%%%%%%%%%%%%%%%%%%%%%%%
%%%%%%%% TABLE 1
%%%%%%%%%%%%%%%%%%%%%%%%%%%%%%%%%%%
\begin{table}
\caption{Redshift and photon indices of the GRB sample$^{\rm a}$.}
\label{tab:grbsample}
\centering
\begin{tabular}{l c c c c c c}
\hline
GRB & Redshift & Photon Index  \\
    & $z$    &    $\Gamma$     \\
\hline
091018  & 0.971 & $2.0   \pm 0.115$       \\
070318  & 0.84  & $2.01  \pm 0.12$ \\
080430  & 0.767 & $1.98  \pm 0.09$ \\
090618  & 0.54  & $1.83  \pm 0.04$ \\
110715A & 0.82  & $1.760 \pm 0.105$ \\
140206A & 2.73  & $1.80  \pm 0.05$ \\
\hline
\end{tabular}
\begin{tablenotes}
\small
\item $^{\rm a}$Redshifts and photon indices are obtained 
from the Swift/\textit{XRT} GRB light curve repository \citep{eva+07,eva+09}.
\end{tablenotes}
\end{table}

%%%%%%%%%%%%%%%%%%%%%%%%%%%%%%%%%%%
%%%%%%%% TABLE 2
%%%%%%%%%%%%%%%%%%%%%%%%%%%%%%%%%%%
\begin{table*}
\caption{Estimated values of the putative nascent magnetar parameters for the changing magnetic dipole moment case.}
\label{tab:estpars}
\begin{tabular}{l c c c c c c}
\hline
GRB & $P_{0}$ & $\sin\alpha_{0}$ & $\mu_{0}$ & $\mu_{\infty}$ & $t_{\rm m}$ & $\chi^{2}/{\rm dof}$ \\
    & (ms)    &                  & ($10^{33}$~G\,cm$^{3}$) & ($10^{33}$~G\,cm$^{3}$) & (days) & \\
\hline
091018  & $2.917 \pm 0.041$ & $0.488 \pm 0.221$ & $1.829 \pm 0.167$ & $0.689 \pm 0.063$ & $0.015 \pm 0.002$ & $139.35/134$  \\
\hline
091018$^{\rm a}$  & $2.922 \pm 0.039$ & $0.523 \pm 0.236$ & $1.804 \pm 0.170$ & $0.683 \pm 0.063$ & $0.016 \pm 0.002$ & $139.23/134$ \\
\hline
070318  & $3.848 \pm 0.064$ & $0.454 \pm 0.226$ & $1.382 \pm 0.127$ & $0.369 \pm 0.038$ & $0.024 \pm 0.002$ & $126.90/90$ \\
080430  & $4.237 \pm 0.070$ & $0.547 \pm 0.200$ & $0.442 \pm 0.043$ & $0.233 \pm 0.030$ & $0.158 \pm 0.057$ & $140.42/140$ \\
090618  & $2.570 \pm 0.014$ & $0.709 \pm 0.082$ & $0.653 \pm 0.028$ & $0.374 \pm 0.019$ & $0.068 \pm 0.008$ & $1188.64/975$ \\
110715A & $1.913 \pm 0.021$ & $0.408 \pm 0.211$ & $0.860 \pm 0.069$ & $0.179 \pm 0.016$ & $0.023 \pm 0.001$ & $362.53/248$ \\
140206A & $0.670 \pm 0.008$ & $0.628 \pm 0.243$ & $0.264 \pm 0.030$ & $0.119 \pm 0.014$ & $0.037 \pm 0.002$ & $594.17/479$ \\
\hline
\end{tabular}
\begin{tablenotes}
\small
\item $^{\rm a}$ Parameter values obtained from numerical analysis presented in \paper.
\end{tablenotes}
\end{table*}

%%%%%%%%%%%%%%%%%%%%%%%%%%%%%%%%%%%%%
\subsection{Approximate analytical solutions of the angular velocity and inclination angle}
\label{sec:appsol}
%%%%%%%%%%%%%%%%%%%%%%%%%%%%%%%%%%%%%

In \paper~we have solved the above set of equations numerically to find the evolution of $\Omega$, $\alpha$ and thus $L_{\rm X}$. Although a single numerical solution takes less than a second, the Bayesian fitting procedure coupled with the MCMC simulation, requires solving the ODE set several hundred thousand times which is computationally expensive.
An exact solution for Equations (\ref{eq:mdr1}) and (\ref{eq:mdr2}) is given in \citet{phi+14a}, but as this solution is implicit, employing it would require solving the algebraic equation numerically at each time step, therefore, using this method does not improve the computational expense. 

We, thus, present a very accurate approximate solution of the angular velocity and inclination angle, i.e.\ the ODE set in Section \ref{sec:modeleq}.
This significantly  (by $\sim$5 times)  reduces the computational time and gives insight into the dependencies of the spin and inclination angle.

Equations~(\ref{eq:mdr1})-(\ref{eq:mdr2}) implies an integration constant
%%%%%%%%%%%%%%%%
\begin{equation}
\Omega\frac{1-\sin^2\alpha}{\sin\alpha}=\Omega_0\frac{1-\sin^2\alpha_0}{\sin\alpha_0}, \label{eq:spin_angle_const} 
\end{equation}
%%%%%%%%%%%%%%%%
where $\Omega_0$ and $\alpha_0$ are the initial values of the spin and the inclination angle, respectively \citep{phi+14a}. By using the integration constant, the angular velocity can be eliminated from Equation (\ref{eq:mdr2}),
%%%%%%%%%%%%%%%%
\begin{equation}
\frac{\mathrm{d} y}{\mathrm{d}\tau}=-\frac{y^3}{1-y^2} \frac{(1-y_0^2)^2}{y_0^2}, 
\label{eq:dif_eq_y}
\end{equation}
%%%%%%%%%%%%%%%%
where $y=\sin\alpha$, $y_0=\sin\alpha_0$ and the dimensionless time, $\tau$, is defined as
%%%%%%%%%%%%%%%%
\begin{equation}
\tau\equiv \frac{\Omega_0^2}{Ic^3} \int^t_0 \mu^2 \,\dif t. 
\label{eq:t_tau}
\end{equation}
%%%%%%%%%%%%%%%%
Integrating Equation~(\ref{eq:dif_eq_y}) gives
%%%%%%%%%%%%%%%%
\begin{equation}
-y_0^2\ln\left(1+\xi\right)+\xi=2\left(1-y_0^2\right)^2\tau, 
\label{eq:impl_sol}
\end{equation}
%%%%%%%%%%%%%%%%
where $\xi=y_0^2/y^2-1$. By applying the approximate solution,
%%%%%%%%%%%%%%%%
\begin{equation}
\xi=\xi_0+y_0^2\xi_1+y_0^4\xi_2,
\end{equation}
%%%%%%%%%%%%%%%%
into Equation (\ref{eq:impl_sol}) and then solving it to the order of $y_0$, an approximate solution for the inclination angle is obtained as
%%%%%%%%%%%%%%%%
\begin{equation}
y\left(\tau\right) =\frac{y_0}{\sqrt{1+2(1-y_0^2)^2\tau+y_0^2\ln\left(1+2\tau\right)+y_0^4\frac{\ln\left(1+2\tau\right)-4\tau}{1+2\tau}}} \, .
\label{eq:app_sol1}
\end{equation}
%%%%%%%%%%%%%%%%
If $y_0$ is zero or one, Equation (\ref{eq:dif_eq_y}) implies trivially $y(\tau)=y_0$. Equation (\ref{eq:app_sol1}) yields the former case, yet, it does not yield the latter one. This can be fixed by modifying the solution as
%%%%%%%%%%%%%%%%
\begin{align}
y\left(\tau\right) =&\frac{y_0}
{\sqrt{1+f\left(\tau\right)}}, 
\label{eq:app_sol2}
\end{align}
%%%%%%%%%%%%%%%%
where
%%%%%%%%%%%%%%%%
\begin{align}
f\left(\tau\right) = & 2(1-y_0^2)^2\tau+y_0^2\ln\left(1+2\tau\right)+y_0^4\frac{\ln\left(1+2\tau\right)-4\tau}{1+2\tau}
\notag \\
&-y_0^8\left(\ln\left(1+2\tau\right)+ \frac{\ln\left(1+2\tau\right)-4\tau}{1+2\tau}\right).
\end{align}
%%%%%%%%%%%%%%%%
The spin evolution can be easily obtained from Equation (\ref{eq:spin_angle_const})
%%%%%%%%%%%%%%%%
\begin{align}
\Omega\left(\tau\right)&=
\Omega_0\left(1-y_0^2\right)
\frac{\sqrt{1+f\left(\tau\right)}}{1-y_0^2+f\left(\tau\right)}. 
\label{eq:app_sol_w}
\end{align}
%%%%%%%%%%%%%%%%
These approximate solutions are well-consistent with the numerical solutions as shown in  \autoref{fig:app_sol}. The relative difference between the approximate solution and numerical solution is less than $6\%$ for $\alpha_0<70^\circ$. Also, the form of the solutions is not altered in the case of changing magnetic field as the field evolution only modifies the relation between the time, $t$, and the dimensionless time, $\tau$ given in Equation (\ref{eq:t_tau}).  

The linear term of $\tau$ increases faster than the logarithmic term $\ln(1+\tau)$. So, in the limit of $\tau \gg 1$, the approximate solution of the inclination angle reduces to
%%%%%%%%%%%%%%%%
\begin{equation}
y\left(\tau\right) \simeq \frac{y_0}{\sqrt{1+2\left(1-y_0^2\right)^2\tau}}, 
\label{eq:lim1_a}
\end{equation}
%%%%%%%%%%%%%%%%
as well as the spin of the star reduces to
%%%%%%%%%%%%%%%%
\begin{equation}
\Omega\left(\tau\right)\simeq
\Omega_0\frac{1-y_0^2}{\sqrt{1+2\left(1-y_0^2\right)^2\tau}}. 
\label{eq:lim1_w}
\end{equation}
%%%%%%%%%%%%%%%%
For later times, $\tau\gg 1/(1-y_0^2)^2$, both the inclination angle and the spin of the star approximate to
%%%%%%%%%%%%%%%%
\begin{align}
y\left(\tau\right) \simeq&\frac{y_0}{(1-y_0^2)\sqrt{2\tau}}, 
\quad\text{and}\quad
\Omega\left(\tau\right)\simeq\frac{\Omega_0}{\sqrt{2\tau}}. 
\label{eq:lim2}
\end{align}
%%%%%%%%%%%%%%%%
Accordingly, both the spin and the inclination angle decrease with $\tau^{-1/2}$ for the late time. The magnetic field and the rotation axis almost aligned ($\alpha< 11\degr$) in this limit. If the magnetic field is constant, this limit indicates a time-scale
%%%%%%%%%%%%%%%%
\begin{equation}
t_{\text{alignment}}\sim 10^{-1}\frac{y_0^2}{(1-y_0^2)^2}\frac{I_{45}}{\mu_{33}^2}\left(\frac{P_0}{1\,\mathrm{ms}}\right)^2\,\mathrm{day},
\end{equation} 
%%%%%%%%%%%%%%%%
where $I_{45}=I/10^{45}\,\mathrm{g\,cm^2}$ and $\mu_{33}=\mu/10^{33}\mathrm{G\,cm^3}$. Alignment takes longer if the magnetic field decreases with time.

%%%%%%%%%%%%%%%%%%%%%%%%%%%%%%%%%%%%%
\subsection{GRB sample}
\label{sec:grbsamp}
%%%%%%%%%%%%%%%%%%%%%%%%%%%%%%%%%%%%%

Our sample in this work contains GRBs 070318, 080430, 090618, 110715A, 140206A and 091018. We included GRB 091018, a source which is also presented in \paper, in order to compare the numerical and approximate analytical solutions. 

In contrast to \paper, we did not restrict our sample only to GRBs with plateau phases since we now have clue that the magnetic dipole moment might be changing in the first day of a nascent magnetar. Thus, it is possible to model GRB afterglow light curves with steeper evolution.

The unabsorbed flux values, redshifts and photon indices of the GRB sample are obtained from the Swift-XRT GRB light curve repository\footnote{\url{http://www.swift.ac.uk/xrt_curves/}} \citep{eva+07,eva+09} and listed in \autoref{tab:grbsample}. We converted the unabsorbed flux values, $F_{\rm X}$, to luminosity values using
\begin{equation}
L = 4 \pi d_{\rm L}^{2}(z) F_{\rm X} k(z).
\end{equation}
Luminosity distance, $d_{\rm L}(z)$, is calculated in a flat $\Lambda$CDM cosmological model using \texttt{astropy.cosmology} subpackage \citep{astropy18}. The cosmological parameters are taken as $H_{0} = 71 \ {\rm km}\ {\rm s}^{-1} \ {\rm Mpc}^{-1}$ and $\Omega_{\rm M} = 0.27$. The cosmological $k$-correction \citep{blo+01}, $k(z)$, is calculated with $k(z) = (1 + z)^{(\Gamma - 2)}$ using redshift and photon index ($\Gamma$) values listed in  \autoref{tab:grbsample} for each GRB.

%%%%%%%%%%%%%%%%%%%%%%%%%%%%%%%%%%%%%
\subsection{Parameter estimation}
\label{sec:parest}
%%%%%%%%%%%%%%%%%%%%%%%%%%%%%%%%%%%%%

We estimated the period, inclination angle, magnetic dipole moment of nascent magnetars at the start of the plateau phase as well as the value of the magnetic dipole moment which the star settles down and the evolution time-scale of this relaxation by using a Bayesian framework. We have given the details of this analysis in \paper. The light curves of the selected GRBs are modelled with
%%%%%%%%%%%%%%%%
\begin{equation}
L_{\rm X} = \eta\frac{ \mu^2 \Omega^4}{c^3} (1 + \sin^2 \alpha ).
\label{eq:Lx}
\end{equation}
%%%%%%%%%%%%%%%%
Here, $\alpha$ and $\Omega$ are calculated using approximate analytical solutions presented in Section \ref{sec:appsol} by Equations (\ref{eq:app_sol2}) and (\ref{eq:app_sol_w}).

We used Gaussian log-likelihood and uniform prior probability to construct the posterior probability distribution with the same prior probabilities given in \paper~ except for GRB 140206A. For this source we decreased the lower limit of the initial rotation period from $0.7\,{\rm ms}$ to $0.5\,{\rm ms}$ as initial analysis suggested a lower value for this source. Finally, we sampled the posterior probability distribution of the parameters with \texttt{emcee} \citep{emcee13,emceerc218} as described in \paper~in detail and obtained the parameter values from the posterior distributions of each parameters.

%%%%%%%%%%%%%%%%%%%%%%%%%%%%%%%%%%%%%%%%%%%%%%
%%%%%%%%%%%%%%%%%%%%%%%%%%%%%%%%%%%%%%%%%%%%%%
\section{Results}
\label{sec:results}
%%%%%%%%%%%%%%%%%%%%%%%%%%%%%%%%%%%%%%%%%%%%%%
%%%%%%%%%%%%%%%%%%%%%%%%%%%%%%%%%%%%%%%%%%%%%%

\begin{figure*}
\begin{center}
\includegraphics[scale=0.9]{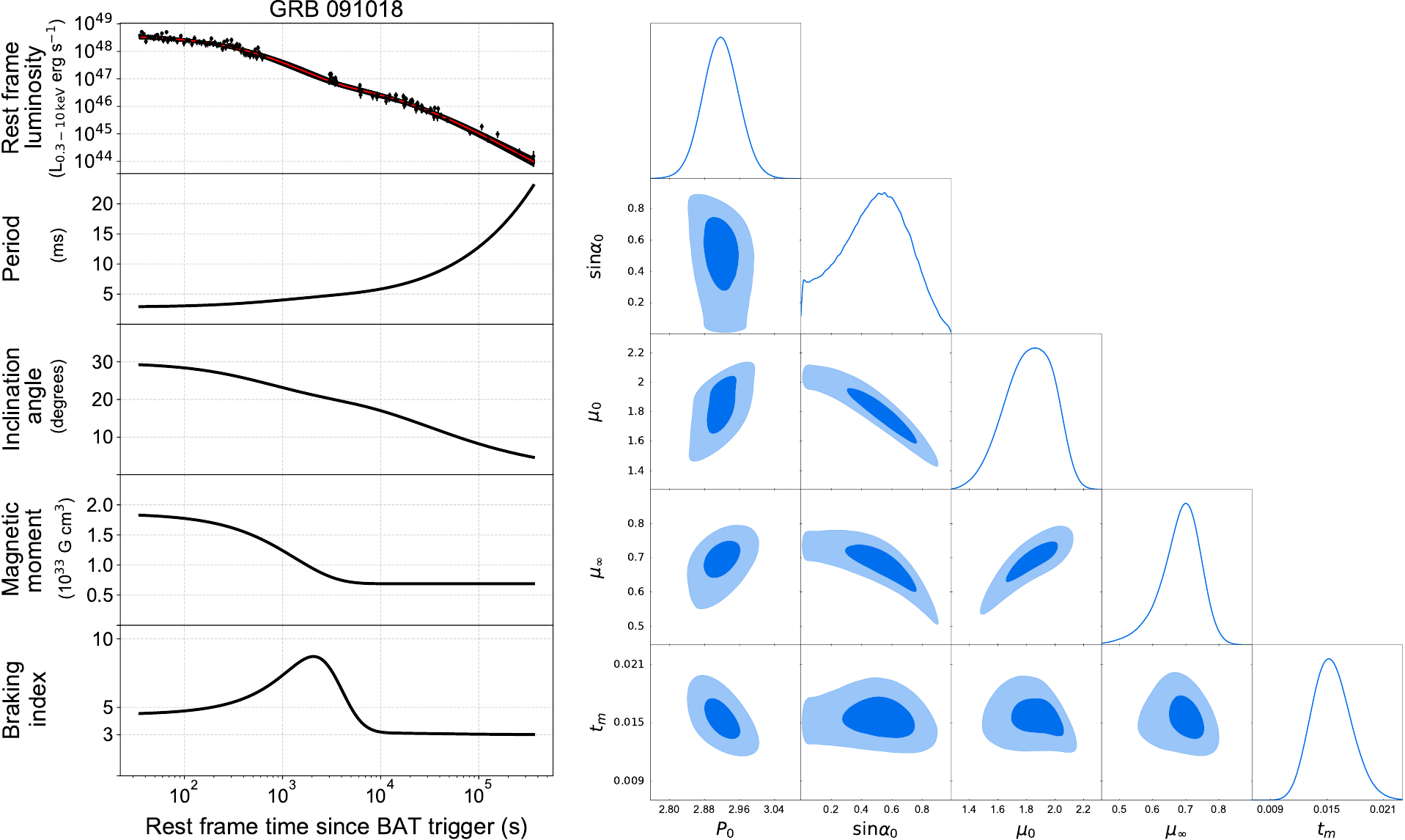}                                  
\caption{{\bf Left:} Evolution of luminosity, period, inclination angle, magnetic dipole moment and braking index of putative nascent magnetar in GRB 091018. The red line in the upper panel represents the luminosity model obtained from the median value of all samples. Solid black lines represent randomly chosen 500 models from the posterior distribution. {\bf Right:} 2D joint (with 1 and 2 sigma contours) and 1D marginalized posterior probability distributions of the parameters plotted with \texttt{getdist} package \citep{getdist18}. \label{fig:091018}}
\end{center}
\end{figure*}

\begin{figure*}
\begin{center}
\includegraphics[scale=0.85]{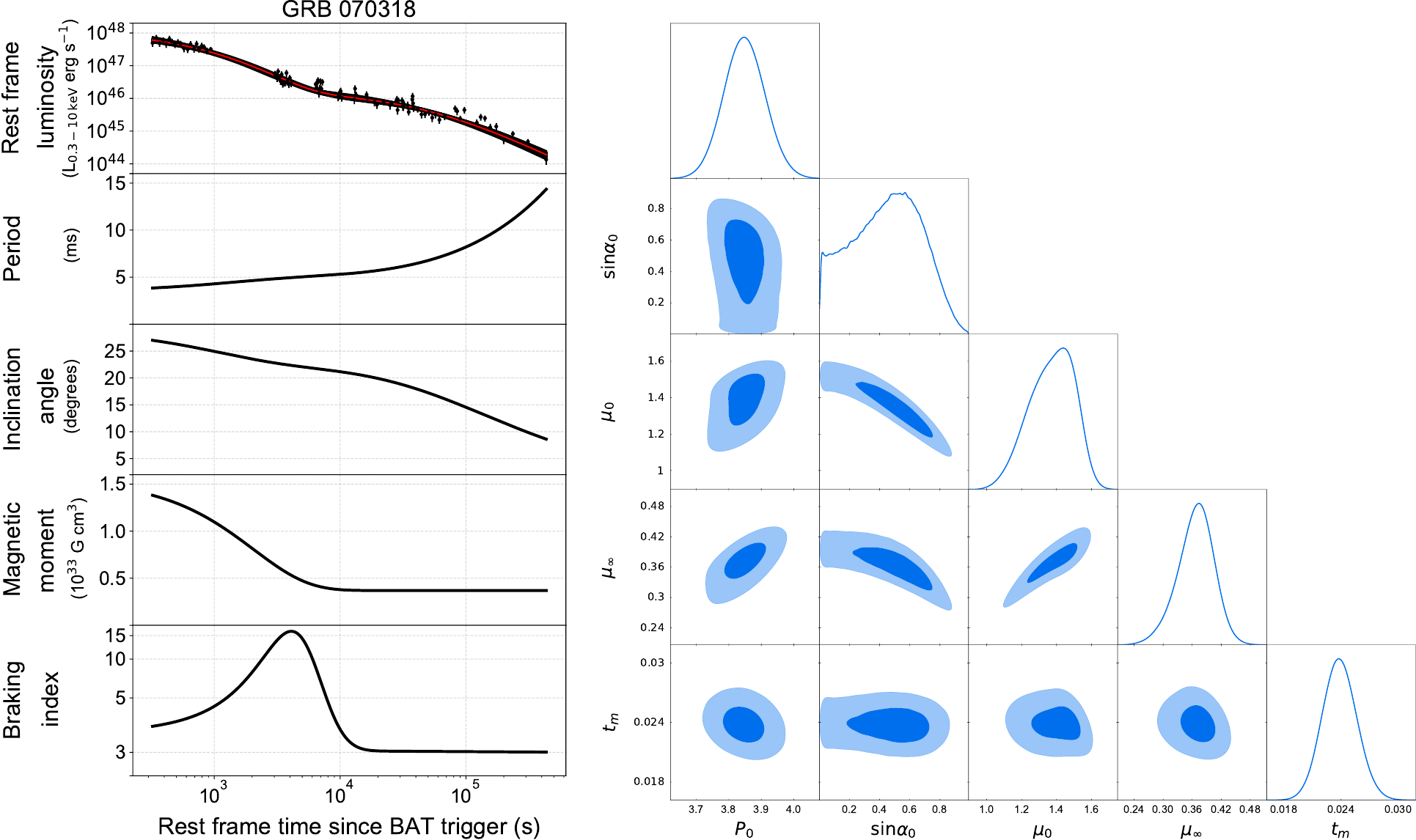}   
\caption{Same as \autoref{fig:091018} but for GRB 070318.\label{fig:070318}}
\end{center}
\end{figure*}

\begin{figure*}
\begin{center}
\includegraphics[scale=0.9]{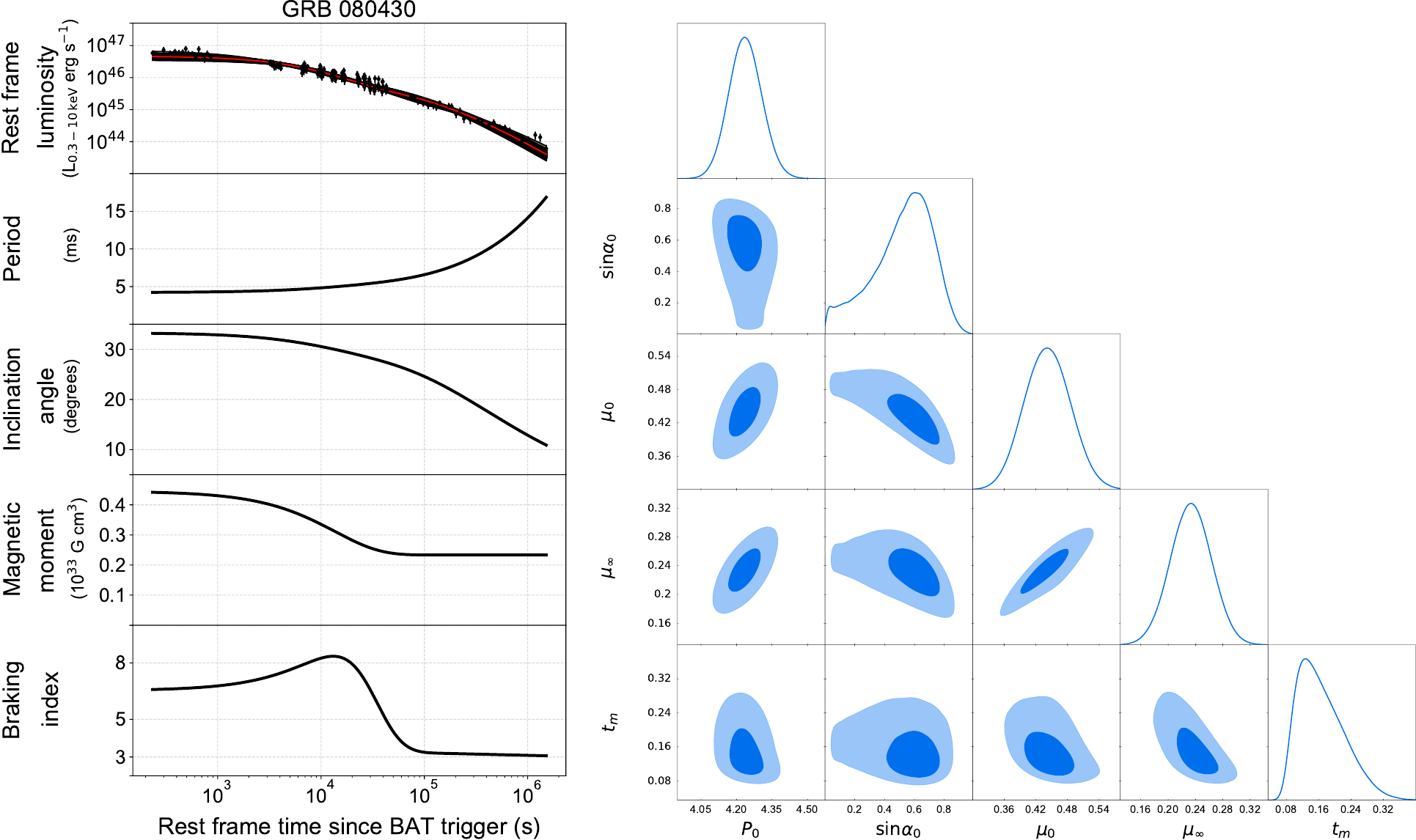}                   
\caption{Same as \autoref{fig:091018} but for GRB 080430.\label{fig:080430}}
\end{center}
\end{figure*}

\begin{figure*}
\begin{center}
\includegraphics[scale=0.9]{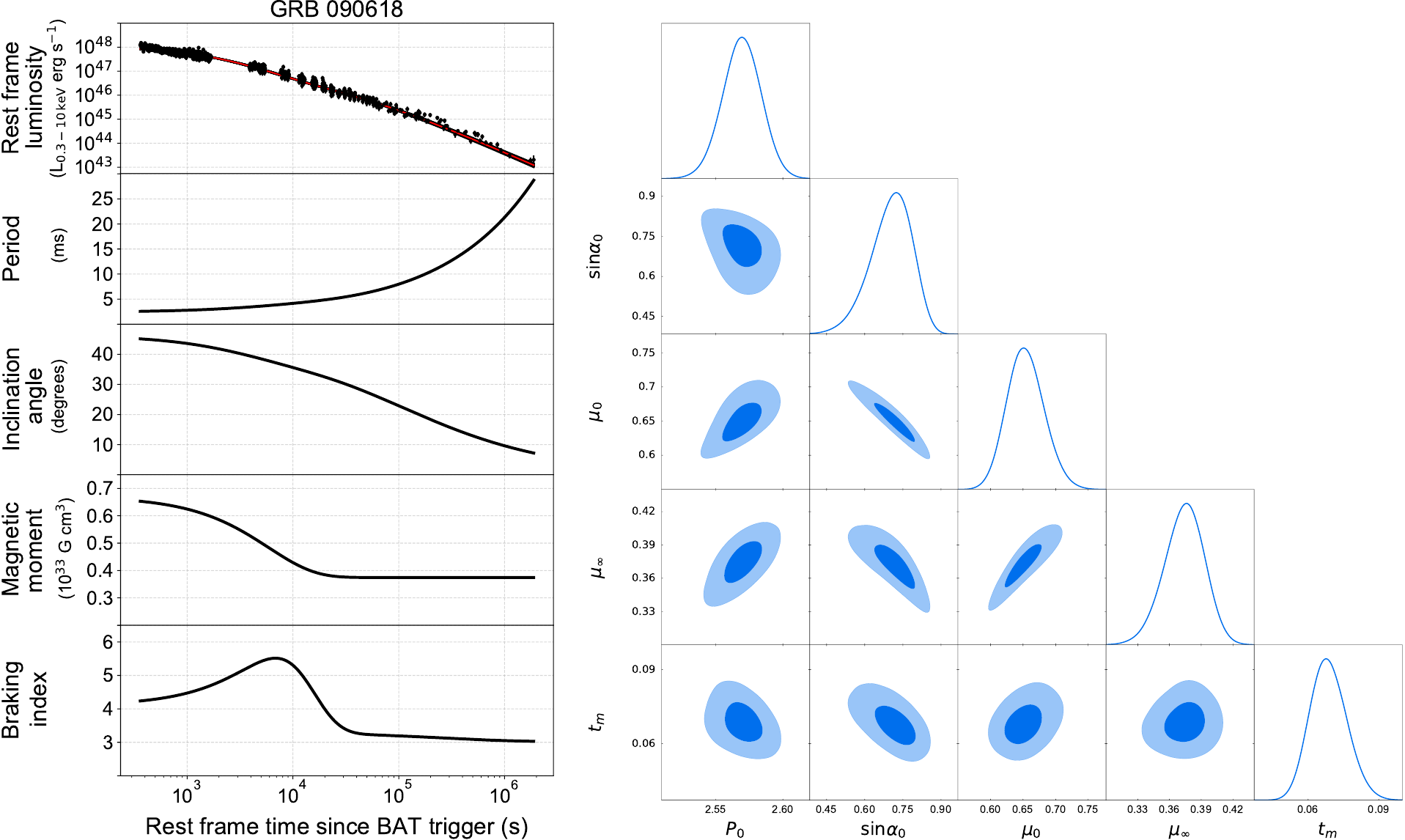}                       
\caption{Same as \autoref{fig:091018} but for GRB 090618.\label{fig:090618}}
\end{center}
\end{figure*}

\begin{figure*}
\begin{center}
\includegraphics[scale=0.9]{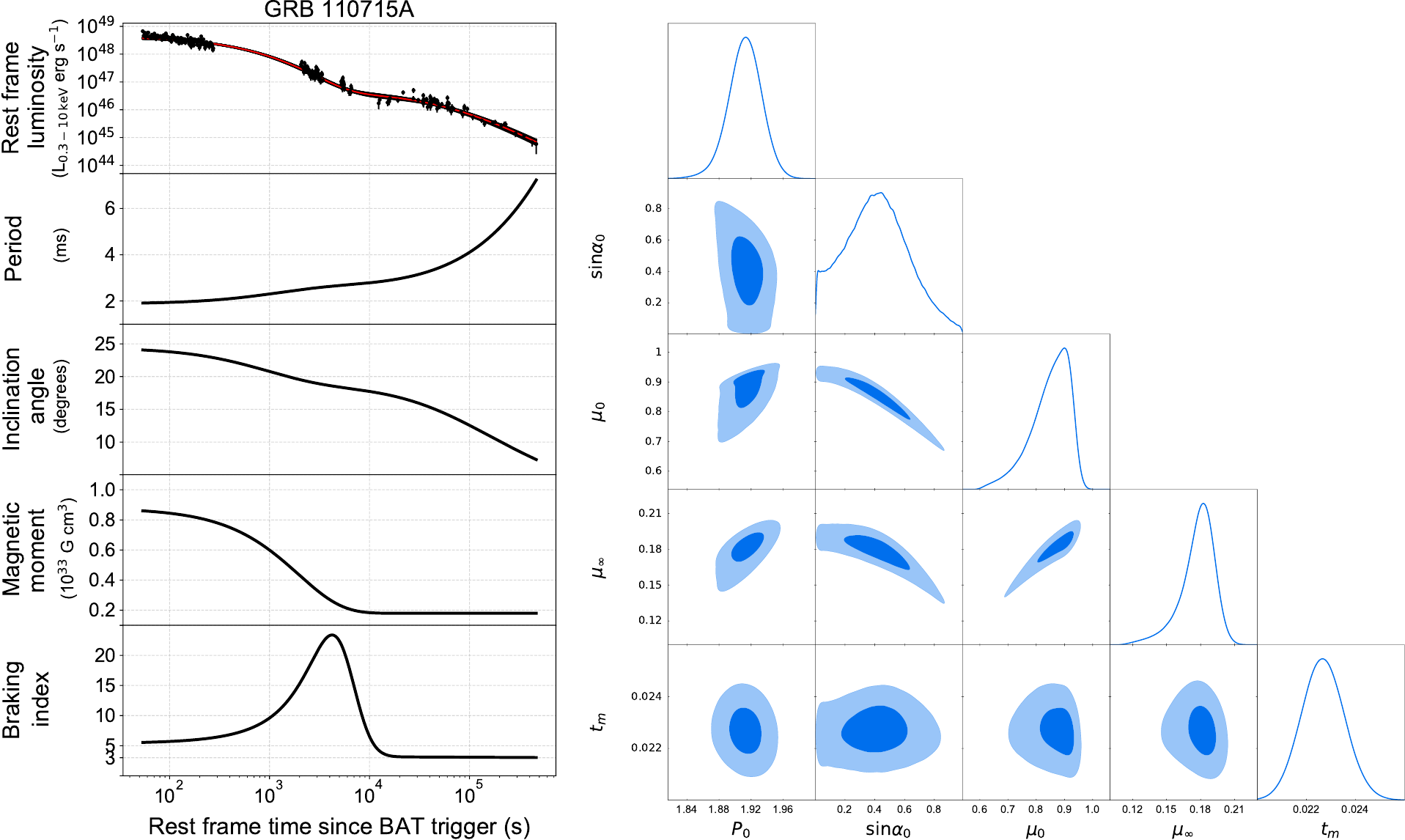}   
\caption{Same as \autoref{fig:091018} but for GRB 110715A.\label{fig:110715a}} 
\end{center}
\end{figure*}

\begin{figure*}
\begin{center}
\includegraphics[scale=0.9]{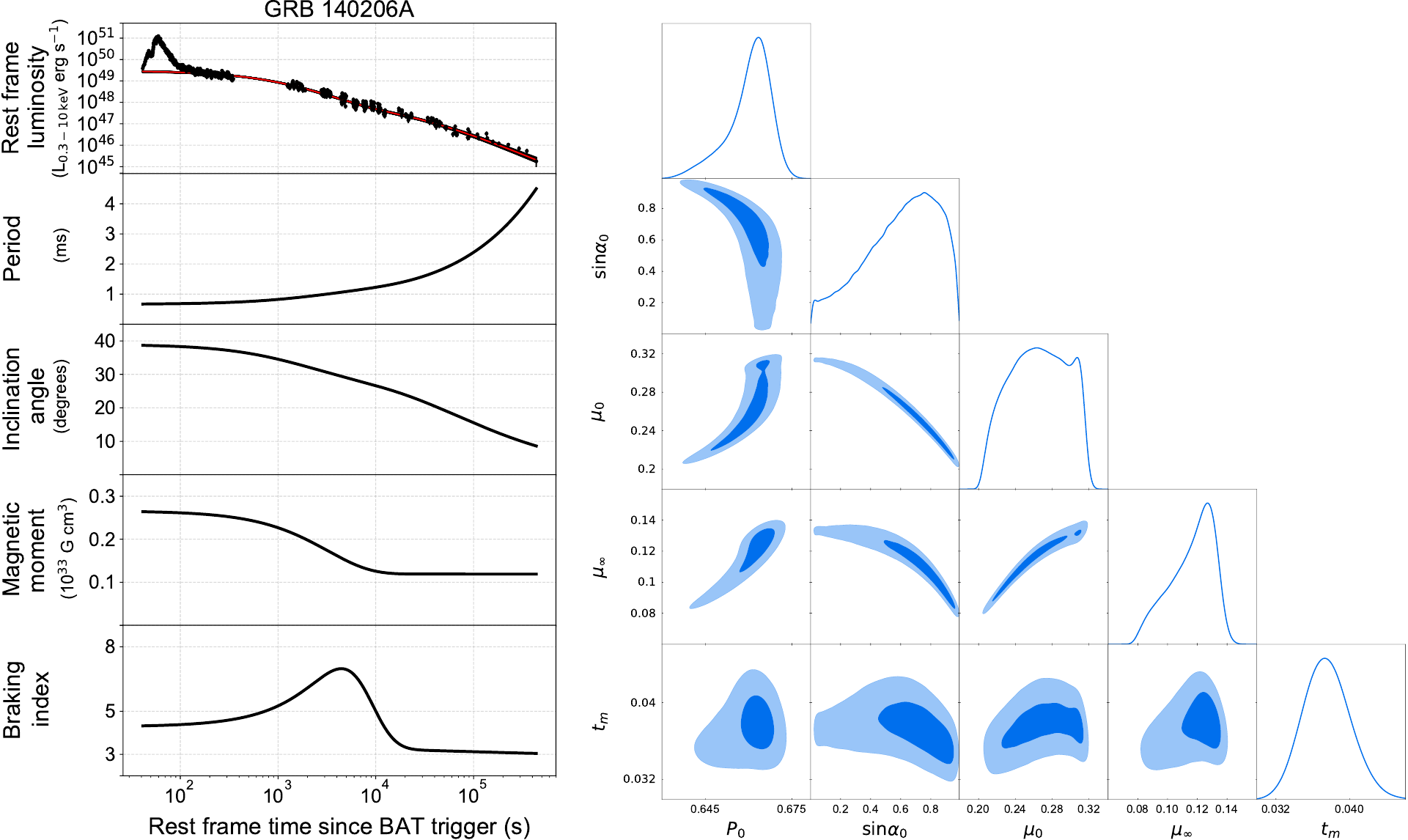}
\caption{Same as \autoref{fig:091018} but for GRB 140206A. We excluded the flare data (coloured in green) that comes after the first data point of the presented light curve.\label{fig:140206a}} 
\end{center}
\end{figure*}

We have modelled the X-ray afterglow light curves of GRB 091018, GRBs 070318, 080430, 090618, 110715A and 140206A with the model described above to determine the initial parameters of the putative magnetars with the Bayesian method introduced above. 
The estimated values of the putative nascent magnetar parameters of the selected GRBs are presented in \autoref{tab:estpars}. The evolution as well as the 1D and 2D posterior distributions of the parameters are presented in \autoref{fig:091018}, \autoref{fig:070318}, \autoref{fig:080430}, \autoref{fig:090618}, \autoref{fig:110715a} and \autoref{fig:140206a}, respectively. We included GRB 091018 in our data set to compare numeric solution presented in \paper~and analytic solution presented in this paper. 

We have found that, within the time frame of the X-ray afterglow---about a few days following the birth of the magnetar---the inclination angle of putative magnetars change from 
$\sim 30^\circ$--$40^\circ$ to $\sim 5^\circ$--$10^\circ$ and the magnetic dipole moments decrease by a factor of 2--5. As a result, the braking index varies significantly in the episodes considered, confirming the previous results of \paper.

\begin{figure*}
\centering
\includegraphics[scale=0.9]{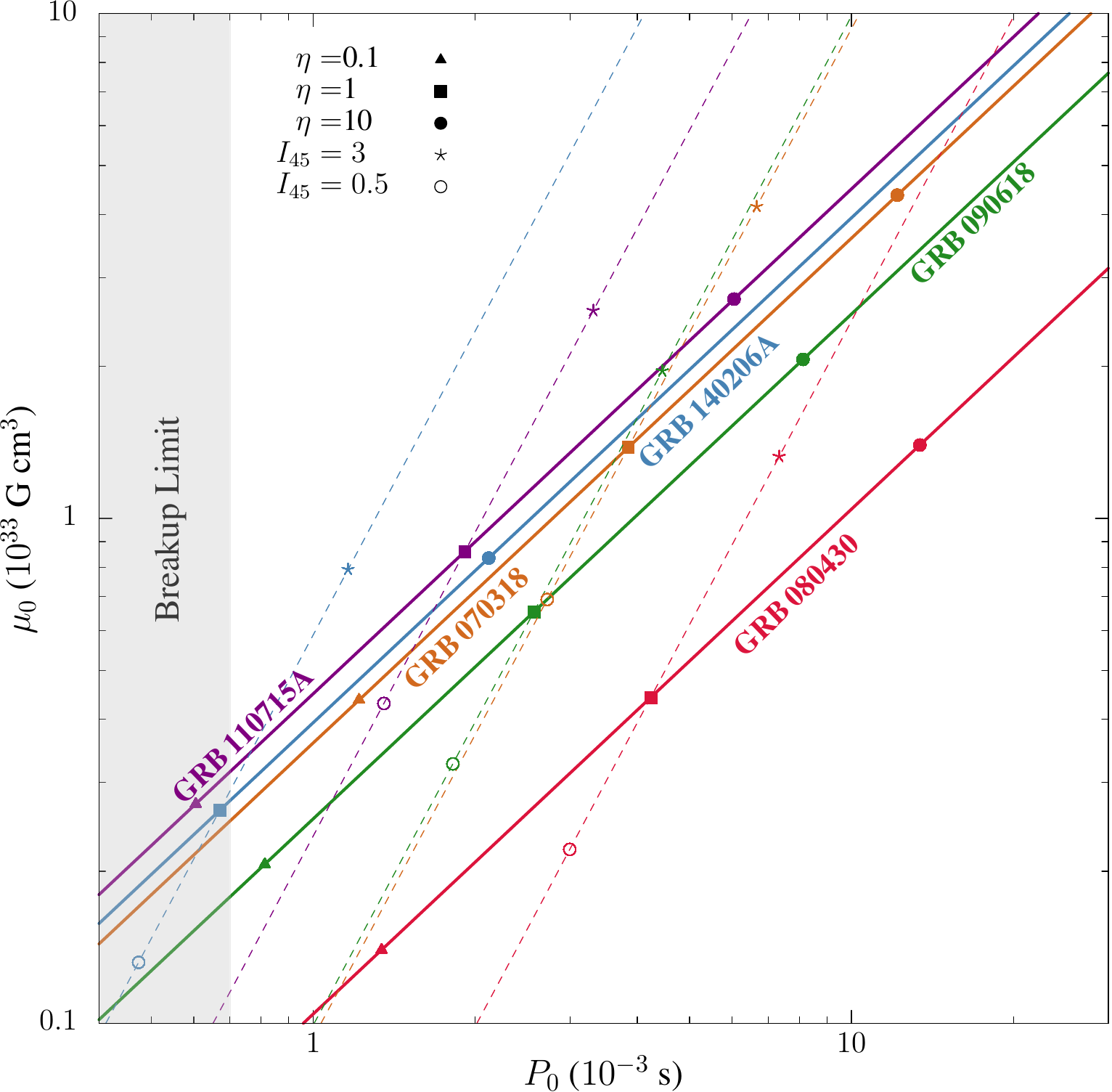}
\caption{Possible values of the initial period and the initial magnetic dipole moment for different values of $\eta$ and $I_{45}$. On solid lines, $\eta$ varies and $I_{45}$ fixed at $1$. On dashed lines, $\eta$ fixed at $1$ and $I_{45}$ varies.  The grey shaded area is the possible minimum period range of various equation of states \citep{coo+94}.\label{fig:muP}}
\end{figure*}

The initial periods and magnetic moments determined in this work depend on the choice of the efficiency factor $\eta$ and moment of inertia $I$. The $\eta$ parameter which involves the X-ray efficiency and the beaming factor varies in a wide range; it can be as low as 10$^{-5}$ or high as 50 \citep{fra+01, Kar+12}. In our simulations we fix $\eta$ as 1, but below, we explain and display in Figure \ref{fig:muP} how the initial parameters transform for different values of $\eta$. The moment of inertia of a neutron star takes values around $10^{45}\,{\rm g\,cm^2}$ depending on the equation of state, the central mass density and the spin of the star \citep{Hae+07}. 
In this work we chose $\eta=1$ and $I_{45}=1$ as is usual to choose.
We note that, $\eta$ and $I$ can be eliminated from the equations by defining new variables as $\sqrt{\eta I_{45}}\Omega$ and $\mu/(I_{45}\sqrt{\eta})$. Therefore, for different values of $\eta$ and $I$, the initial parameters  transform as $P_0\rightarrow P_0\sqrt{\eta I_{45}}$ and $\mu_0\rightarrow \mu_0 I_{45}\sqrt{\eta}$ while the others do not change. In \autoref{fig:muP} we present all possible values for each source on the $\mu_0-P_0$ plane. We emphasize that the evolution of the inclination angle and the braking index are not affected by the choice of $\eta$ or $I$. 

Recently, \cite{xia17, xia19} calculated the X-ray efficiency factor as a function of luminosity based on an emission mechanism governed by Poynting flux-dominated wind. This implies that the value of $\eta$ may not be constant during the episodes we consider. Yet given that $\eta$ depends also on the beaming fraction, employing any possible dependence on the luminosity will not improve our estimates on the initial parameters. Considering the dependence of $\eta$ on luminosity and beaming will be carried over in a future work and is beyond the scope of the present paper.

%%%%%%%%%%%%%%%%%%%%%%%%%%%%%%%%%%%%%%%%%%%%%%
\section{Summary and discussion}
\label{sec:discuss}
%%%%%%%%%%%%%%%%%%%%%%%%%%%%%%%%%%%%%%%%%%%%%%
%%%%%%%%%%%%%%%%%%%%%%%%%%%%%%%%%%%%%%%%%%%%%%
We have invoked the `millisecond magnetar model' \citep{uso92,dun92} to infer the initial parameters of nascent magnetars from the X-ray afterglow light curves of GRBs. 

We have presented an explicit approximate analytical solution of the system of equations describing the evolution of spin and inclination angle of a magnetized neutron star. 
We have shown that this solution is very accurate except for highly orthogonal initial conditions ($\alpha_0> 70^\circ$). 

We have fitted, via a Bayesian procedure, the light curves of 6 GRB afterglows by using the analytical solution to determine the evolution of the period, inclination angle, magnetic dipole moment and the braking index. The spin and magnetic parameters we obtained are consistent with the initial parameters suggested for the `millisecond magnetar model'.

We have shown that the inclination angle, just like the spin period, varies rapidly within the time-frame of the X-ray afterglows. This is compatible with the recent result we have obtained that the inclination angles of magnetars align rapidly within the first $\sim 10$~days (\paper). As a consequence of the alignment and magnetic field decay the braking index is greater than three ($n>3$) and varies rapidly confirming the results of (\paper). According to this picture the constant braking indices inferred by \citet{las+17} and \citet{lu+19} are effective average values.

Magnetohydrodynamics simulations employed for nascent magnetars have shown that these stars continue their lifes with magnetar strength magnetic fields if the star has a high rotation period (P $<$ 6 ms) and small inclination angle ($\alpha < 45^\circ$) \citep{gep+06}. Although we can not give a limit on period due to its dependence on the poorly constrained $\eta$ parameter, all inclination angle values in our sample are smaller than $45^\circ$ i.e.\ compatible with the theoretical prediction of \citet{gep+06}. We thank Prof. Geppert for bringing into our attention this interesting prediction.

The `millisecond magnetar model' is often invoked as an explanation to the `plateau phase' observed in some X-ray afterglows. We have shown that the magnetic field of a nascent magnetar may decline immediately after its birth. 
Most of the models  in the literature (e.g.\ \cite{col+00}) consider the long-term evolution of magnetic fields of magnetars with solid crusts. This is a quasi-equilibrium stage. Evolutionary time-scales observed in these simulations are hundreds of years. The brief episode we consider in this paper is very soon after the initial generation and enhancement of the field by magnetohydrodynamics instabilities where the quasi-equilibrium stage has not yet been achieved and the field may decay more rapidly \citep{gep+06, ben+17}.
As a result the spin-down power of the magnetar decreases more rapidly than it would if magnetic field remained constant, and thus an extended  `plateau phase' may not be realized. According to this picture the systems with the extended `plateau phase' host magnetars with relatively longer field decay time-scales. This suggests that the relevance of the `millisecond magnetar model' may not be restricted to the GRB afterglows with a plateau phase. 

\section*{Acknowledgements}

This work made use of data supplied by the UK Swift Science Data Centre at the University of Leicester (http://www.swift.ac.uk/xrt{\_}curves/). S\c{S}M acknowledges post-doctoral research support from \.{I}stanbul Technical University. S\c{S}M and KYE acknowledges support from T{\"U}B{\.I}TAK with grant number 118F028.

%%%%%%%%%%%%%%%%%%%%%%%%%%%%%%%%%%%%%%%%%%%%%%%%%%



\bibliographystyle{mnras}
{\small
\bibliography{refs_ssm} % if your bibtex file is called example.bib
}

%\pagebreak
%\newpage

%\begin{thebibliography}{99}
%\end{thebibliography}

%%%%%%%%%%%%%%%%%%%%%%%%%%%%%%%%%%%
%%%%%%%% FIGURE 2
%%%%%%%%%%%%%%%%%%%%%%%%%%%%%%%%%%%

%%%%%%%%%%%%%%%%%%%%%%%%%%%%%%%%%%%
%%%%%%%% FIGURE 3
%%%%%%%%%%%%%%%%%%%%%%%%%%%%%%%%%%%

%%%%%%%%%%%%%%%%%%%%%%%%%%%%%%%%%%%
%%%%%%%% FIGURE 4
%%%%%%%%%%%%%%%%%%%%%%%%%%%%%%%%%%%

%%%%%%%%%%%%%%%%%%%%%%%%%%%%%%%%%%%
%%%%%%%% FIGURE 5
%%%%%%%%%%%%%%%%%%%%%%%%%%%%%%%%%%%

%%%%%%%%%%%%%%%%%%%%%%%%%%%%%%%%%%%
%%%%%%%% FIGURE 6
%%%%%%%%%%%%%%%%%%%%%%%%%%%%%%%%%%%

%%%%%%%%%%%%%%%%%%%%%%%%%%%%%%%%%%%
%%%%%%%% FIGURE 7
%%%%%%%%%%%%%%%%%%%%%%%%%%%%%%%%%%%

%%%%%%%%%%%%%%%%%%%%%%%%%%%%%%%%%%%
%%%%%%%% FIGURE 8
%%%%%%%%%%%%%%%%%%%%%%%%%%%%%%%%%%%

% Don't change these lines
\bsp	% typesetting comment
\label{lastpage}
\end{document}